\documentclass[aps,preprint]{revtex4}
\usepackage{amssymb}
\usepackage{amsmath}
\usepackage{graphicx}
\usepackage{epstopdf}
\begin{document}

\title{{\bf
Magnetized hot neutron matter: lowest order constrained variational
calculations}}

\author{{\bf G.H. Bordbar $^{1,2}$ \footnote{Corresponding author. E-mail:
bordbar@physics.susc.ac.ir}}, {\bf Z. Rezaei$^{1}$}}
 \affiliation{ $^1$Department of Physics,
Shiraz University,
Shiraz 71454, Iran\footnote{Permanent address},\\
 and\\
$^2$Research Institute for Astronomy and Astrophysics of Maragha,
P.O. Box 55134-441, Maragha 55177-36698, Iran}

%%%%%%%%%%%%%%%%%%%%%%%%%%%%%%%%%%%%%%%%%%%%%%%%%%%%%%%%%%%%%%%%%%%%%%%%%

\begin{abstract}
We have studied the spin polarized hot neutron matter in the
presence of strong magnetic field. In this work, using the lowest
order constrained variational method at finite temperature and
employing $AV_{18}$ nuclear potential, some thermodynamic properties
of spin polarized neutron matter such as spin polarization
parameter, free energy, equation of state and effective mass have
been calculated. It has been shown that the strong magnetic field
breaks the symmetry of the free energy, leading to a magnetized
equilibrium state. We have found that the equation of state becomes
stiffer by increasing both magnetic field and temperature. The
magnetic field dependence of effective mass for the spin-up and
spin-down neutrons has been investigated.
\end{abstract}

\keywords{ Neutron matter - strong magnetic field - equation of
state}

% \pacs{21.65.-f, 26.60.-c, 64.70.-p}

\maketitle
%%%%%%%%%%%%%%%%%%%%%%%%%%%%%%%%%%%%%%%%%%%%%%%%%%%%%%%

\section{INTRODUCTION}

Based on the supernova models, after the gravitational collapse
of a degenerate stellar core and the explosive ejection of outer
layers material, a protoneutron star would be born \cite{Haensel}.
Through the formation of the protoneutron star, the system arrives
at a temperature of about $20-50\ MeV$ \cite{Camenzind}. This
protoneutron star is hot, opaque to neutrinos, and larger than an
ordinary neutron star \cite{Haensel}. After the formation of the
protoneutron star, neutrino emission is the dominant process in
cooling of the neutron star (mainly by URCA process and neutrino
Bremsstrahlung) \cite{Camenzind}.

Woltjer has predicted a magnetic field strength of order $10^{15}\
G$ for neutron stars as a result of the magnetic flux conservation
from the progenitor star \cite{Woltjer}. This in agreement with the
experimental indication that the surface magnetic field strength of
magnetars can be of the order $B_{magnetar}\approx 10^{14}-10^{15}\
G$ \cite{Thompson,Lazzati}. The magnetic field can be distorted or
amplified by some mixture of convection, differential rotation, and
magnetic instabilities \cite{Tayler,Spruit}.
The relative importance of these ingredients depends on the initial
field strength and the rotation rate of the star. For both
convection and differential rotation, the field and its supporting
currents are not likely to be confined to the solid crust of the
star but instead distributed in most of the stellar interior, which
is mostly a fluid mixture of neutrons, protons, electrons, and other
more exotic particles \cite{Reisen}. Thompson et al. argued that
newborn neutron stars probably combine vigorous convection and
differential rotation, making it likely that a dynamo process might
operate in them \cite{Thompson2}. They expected fields up to
$10^{15}-10^{16}\ G$ in neutron stars with few-millisecond initial
periods.
In the core of high density inhomogeneous gravitationally bound
neutron stars, the magnetic field strength can be as large as
$10^{20}\ G$ \cite{Ferrer}.
In addition, considering the formation of a quark core in the high
density interior of a neutron star, the maximum field reaches up to
about $10^{20}\ G$ \cite{Ferrer,Tatsumi}.
According to the scalar virial theorem based on Newtonian gravity,
the magnetic field strength is allowed to be up to $10^{18}\ G$ in
the interior of a magnetar \cite{Lai}. Moreover, general relativity
predicts the allowed maximum value of the neutron star magnetic
field to be $10^{18}-10^{20}\ G$ \cite{shapiro}. By comparing the
cooling curves of neutron stars with the observational data, Yuan et
al. obtained the magnetic field strength of order $10^{19}\ G$ for
many not so old neutron stars \cite{Yuan}.

The finite temperature and strong magnetic field in the interior of
a protoneutron star can influence different astrophysical
quantities. Therefore, to have a better understanding of different
astrophysical phenomena such as supernova explosion, thermal
evolution and cooling of protoneutron stars, gravitational wave
emission spectrum from neutron star mergers, and to get the more
precise astrophysical quantities such as properties of very young
hot neutron stars and composition of neutron stars, one should
consider the neutron star matter at finite temperatures (excited
neutron star matter) and strong magnetic fields.

Since $\beta$-equilibrium leads to an increase in the number of
neutrons in neutron star matter, it is possible to approximate the
neutron star matter by the pure neutron matter.
Many works have dealt with the study of dense neutron matter at
finite temperature  \cite{Alonso2,
Mukherjee,Panda,Rios2,Bombaci,Rios3}.
Alonso et al. have used a field theoretical model for the analysis
of relativistic neutron matter \cite{Alonso2}. By solving the model
in the renormalized Hartree approximation, they have investigated
the effect of central temperature on the maxima of mass for stable
configurations, the radii of the configurations, and the
gravitational red shift at the surface of a neutron star.
Panda et al. used a mean-field description of nonoverlaping nucleon
bags bound by the self-consistent exchange of $\sigma$, $\omega$,
and $\rho$ mesons to investigate the properties of neutron matter at
finite temperature \cite{Panda}. They showed that by increasing the
temperature, the equation of state becomes stiffer.
Within the framework of the Brueckner-Hartree-Fock formalism and
using the $AV_{18}$ nucleon-nucleon interaction for the spin
polarized neutron matter, Bombaci et al. found that an increase in
the temperature moderately affects the single-particle potentials
\cite{Bombaci}.
In the Hartree-Fock approximation using Skyrme type interactions for
spin polarized neutron matter, Rios et al. showed that the critical
density at which the ferromagnetism takes place, decreases by
temperature \cite{Rios2}.
Within the self-consistent Green's-function approach applying the CD
Bonn and the $AV_{18}$ potential for neutron matter, Rios et al.
found that the effect of dynamical correlations on the macroscopic
properties is rather insensitive to the thermal effects
\cite{Rios3}.
The variational theory for fermions at finite temperature and high
density has been applied by Mukherjee to neutron matter
\cite{Mukherjee}. It has been found that the temperature dependence
of the correlation operator is weak, but it is not negligible.
Besides, it has been shown that the first order phase transition due
to neutral pion condensation has a critical temperature of about $22
MeV$ for neutron matter.
The effect of strong magnetic field on the properties of dense
neutron matter has also been considered.
Perez-Garcia showed that for the neutron matter in the presence of
strong magnetic fields, in the Skyrme model, there is a
ferromagnetic phase transition at $\rho \sim 4\rho_{0}$, whereas it
is forbidden in the $D1P$ model \cite{Perez-Garcia}. The results
indicate that the effects of temperature on the neutron
magnetization remain moderate at temperatures up to about $T = 40\
MeV$.
In the context of the Landau theory of normal Fermi liquids, using
Skyrme and Gogny effective interactions, some thermodynamical
quantities such as isothermal compressibility and spin
susceptibility of pure neutron matter have also been studied
\cite{Perez-Garcia2}.

In our previous works, we have investigated the spin polarized
neutron matter \cite{Bordbar75}, symmetric nuclear matter
\cite{Bordbar76}, asymmetric nuclear matter as well as neutron star
matter \cite{Bordbar77} and magnetized neutron matter
\cite{Bordbar83} at zero temperature using lowest order constrained
variational (LOCV) method with the realistic strong interactions. We
have also investigated the thermodynamic properties of spin
polarized neutron matter \cite{Bordbar78}, symmetric nuclear matter
\cite{Bordbar80}, and asymmetric nuclear matter \cite{Bordbar81} at
finite temperature with no magnetic field.
In the present work, we calculate different thermodynamic properties
of spin polarized neutron matter at finite temperature in the
presence of strong magnetic field using LOCV technique employing
$AV_{18}$ potential.

%-----------------------------------------------------------

\section{LOCV formalism for the spin polarized hot neutron matter in the
presence of magnetic field} We consider a homogeneous system of $N$
interacting particles with $N^{(+)}$ spin-up and $N^{(-)}$ spin-down
neutrons under the influence of a uniform magnetic field, i.e.
$\mathbf{B}=B\widehat{k}$. The number densities of spin-up and
spin-down neutrons are presented by $\rho^{(+)}$ and $\rho^{(-)}$
respectively. We introduce the spin polarization parameter,
$\delta$, by
\begin{eqnarray}
     \delta=\frac{\rho^{(+)}-\rho^{(-)}}{\rho},
 \end{eqnarray}
where $-1\leq\delta\leq1$, and $\rho$ is the total number density of
system. The magnetization density of neutron matter is defined as
\begin{eqnarray}\label{m1}
     m=\mu_n \delta \rho ,
 \end{eqnarray}
where $\mu_n$ is the neutron magnetic moment. The total
magnetization of a given volume is also as follows
\begin{eqnarray}\label{m2}
     M= \int m dV.
 \end{eqnarray}

In order to calculate the energy of this system, we use LOCV method
as follows: we consider a trial many-body wave function of the form
\begin{eqnarray}
     \psi=\mathcal{F}\phi,
 \end{eqnarray}
where $\phi$  is the uncorrelated ground-state wave function of $N$
independent neutrons, and $\mathcal{F}$ is a proper $N$-body
correlation function. Using Jastrow approximation \cite{Jastrow},
$\mathcal{F}$ can be replaced by
\begin{eqnarray}
    \mathcal{F}=S\prod _{i>j}f(ij),
 \end{eqnarray}
where $S$ is a symmetrizing operator. We consider a cluster
expansion of the energy functional up to the two-body term,
 \begin{eqnarray}\label{tener}
           E([f])=\frac{1}{N}\frac{\langle\psi|H|\psi\rangle}
           {\langle\psi|\psi\rangle}=E _{1}+E _{2}\cdot
 \end{eqnarray}
The one-body term, $E_{1}$, is given
by
 \begin{eqnarray}\label{E1}
 E_{1}=-\frac{M H}{N}+\sum_{i=+,-}\overline{\varepsilon}_{i},
\end{eqnarray}
where the first term of Eq. (\ref{E1}), $H$ shows the external magnetic
field. It should be noted that we have used $B$ to present the
magnetic field strength; while the total magnetic field is the sum
of the external magnetic field and the induced magnetization, $B=H+4
\pi M$. Because of the tiny value of the neutron magnetic moment, we
assume that the induced magnetization has a small contribution to
the total magnetic field. Consequently,
\begin{eqnarray}\label{BH}
B\sim H.
\end{eqnarray}
Using Eqs. (\ref{m1}), (\ref{m2}) and (\ref{BH}), the one-body term
can be written as
 \begin{eqnarray}
 E_{1}=-\mu_n B \delta+\sum_{i=+,-}\overline{\varepsilon}_{i}.
\end{eqnarray}
The second term in Eq. (\ref{E1}), $\overline{\varepsilon}_{i}$ is
as follows
 \begin{eqnarray}
\overline{\varepsilon}_{i}=\sum_{k}\frac{\hbar^{2}k^{2}}{2m}
\overline{n}_{i}(k,T,B,\rho^{(i)}),
\end{eqnarray}
 where $\overline{n}_{i}(k,T,B,\rho^{(i)})$ is the Fermi-Dirac
distribution function in the presence of magnetic field,
 \begin{eqnarray}
\overline{n}_{i}(k,T,B,\rho^{(i)})=\frac{1}{e^{\beta[\overline{\epsilon}_{i}
(k,T,B,\rho^{(i)})-\overline{\mu}_{i}(T,B,\rho^{(i)})]}+1}.
\label{FDB}
\end{eqnarray}
In Eq. (\ref{FDB}), $\overline{\epsilon}_{i}$ and
$\overline{\mu}_{i}$ are the single-particle energy of a neutron and
the neutron chemical potential respectively.
The single-particle energy, $\overline{\epsilon}_{i}$, of a neutron
with momentum $k$ and spin projection $i$ in the presence of
magnetic field is approximately written in terms of the effective
mass as follows \cite{Rios2}
\begin{eqnarray}
\overline{\epsilon}_{i}(k,T,B,\rho^{(i)})&=&
\left\{\begin{array}{ll}
\frac{\hbar^{2}k^{2}}{2m_{+}^{*}(T,\rho)}-\mu_n B+U_{+}(T,\rho^{(+)}) &;~~ i=+,   \\ \\
\frac{\hbar^{2}k^{2}}{2m_{-}^{*}(T,\rho)}+\mu_n
B+U_{-}(T,\rho^{(-)}) &;~~ i=-.
\end{array}
\right.
\end{eqnarray}
In fact, we use a quadratic approximation for the single particle
potential incorporated in the single particle energy as a momentum
independent effective mass. $U_{i}(T,\rho^{(i)})$ is the momentum
independent single particle potential. The effective mass,
$m_{i}^{*}$, is determined variationally
\cite{Modarres3,Modarres5,Modarres7,Modarres8,Friedman}.
The chemical potential, $\overline{\mu}_{i}$, is also obtained by
applying the constraint
 \begin{eqnarray}
\sum_{k}\overline{n}_{i}(k,T,B,\rho^{(i)})=N^{(i)}.
\end{eqnarray}
The two-body energy, $E_{2}$, is
 \begin{eqnarray}
    E_{2}&=&\frac{1}{2N}\sum_{ij} \langle ij\left| \nu(12)\right|
    ij-ji\rangle,
 \end{eqnarray}
where
$$\nu(12)=-\frac{\hbar^{2}}{2m}[f(12),[\nabla
_{12}^{2},f(12)]]+f(12)V(12)f(12).$$  $f(12)$ and $V(12)$ are the
two-body correlation function and nuclear potential respectively. In
the LOCV formalism, the two-body correlation function, $f(12)$, is
considered as follows \cite{Owen},
\begin{eqnarray}\label{correl}
f(12)&=&\sum^3_{k=1}f^{(k)}(r_{12})P^{(k)}_{12},
\end{eqnarray}
where
\begin{eqnarray}\label{Pcor}
P^{(k=1-3)}_{12}&=&(\frac{1}{4}-\frac{1}{4}\sigma_{1}.\sigma_{2}),\
(\frac{1}{2}+\frac{1}{6}\sigma_{1}.\sigma_{2}+\frac{1}{6}S_{12}),\
(\frac{1}{4}+\frac{1}{12}\sigma_{1}.\sigma_{2}-\frac{1}{6}S_{12}).
\end{eqnarray}
In Eq. (\ref{Pcor}), $S_{12}$ and $\sigma_{1}.\sigma_{2}$ are the
tensor and Pauli operators respectively. Using the above two-body
correlation function and the $AV_{18}$ two-body potential
\cite{Wiringa}, after doing some algebra, we find the following
equation for the two-body energy:
\begin{eqnarray}\label{ener2}
    E_{2} &=& \frac{2}{\pi ^{4}\rho }\left( \frac{\hbar^{2}}{2m}\right)
    \sum_{JLSS_{z}}\frac{(2J+1)}{2(2S+1)}[1-(-1)^{L+S+1}]\left| \left\langle
\frac{1}{2}\sigma _{z1}\frac{1}{2}\sigma _{z2}\mid
SS_{z}\right\rangle \right| ^{2}\times  \nonumber
\\&& \times\int dr\left\{\left [{f_{\alpha
}^{(1)^{^{\prime }}}}^{2}{a_{\alpha
}^{(1)}}^{2}(r,\rho,T)\right.\right. \nonumber \\&&\left. \left.
+\frac{2m}{\hbar^{2}}(\{V_{c}-3V_{\sigma } +V_{\tau }-3V_{\sigma
\tau }+2(V_{T}-3V_{\sigma T }) -2V_{\tau z}\}{a_{\alpha
}^{(1)}}^{2}(r,\rho,T)\right.\right. \nonumber \\&&\left.\left.
+[V_{l2}-3V_{l2\sigma } +V_{l2\tau }-3V_{l2\sigma \tau }]{c_{\alpha
}^{(1)}}^{2}(r,\rho,T))(f_{\alpha }^{(1)})^{2}\right ]
+\sum_{k=2,3}\left[ {f_{\alpha }^{(k)^{^{\prime }}}}^{2}{a_{\alpha
}^{(k)}}^{2}(r,\rho,T)\right.\right. \nonumber \\&&\left. \left.
+\frac{2m}{\hbar^{2}}( \{V_{c}+V_{\sigma }+V_{\tau } +V_{\sigma \tau
}+(-6k+14)(V_{t\tau}+V_{t})-(k-1)(V_{ls\tau }+V_{ls})\right.\right.
\nonumber
\\&&\left.\left. +2[V_{T}+V_{\sigma T }+(-6k+14)V_{tT}-V_{\tau
z}]\}{a_{\alpha }^{(k)}}^{2}(r,\rho,T)\right.\right. \nonumber
\\&&\left.\left. +[V_{l2}+V_{l2\sigma } +V_{l2\tau }+V_{l2\sigma \tau
}]{c_{\alpha }^{(k)}}^{2}(r,\rho,T)+[V_{ls2}+V_{ls2\tau }]
{d_{\alpha }^{(k)}}^{2}(r,\rho,T)) {f_{\alpha }^{(k)}}^{2}\right
]\right. \nonumber \\&&\left.
+\frac{2m}{\hbar^{2}}\{V_{ls}+V_{ls\tau }-2(V_{l2}+V_{l2\sigma
}+V_{l2\sigma \tau } +V_{l2\tau })-3(V_{ls2} +V_{ls2\tau
})\}b_{\alpha }^{2}(r,\rho,T)f_{\alpha }^{(2)}f_{\alpha
}^{(3)}\right. \nonumber \\&&\left. +\frac{1}{r^{2}}(f_{\alpha
}^{(2)} -f_{\alpha }^{(3)})^{2}b_{\alpha }^{2}(r,\rho,T)\right\},
 \end{eqnarray}
 where $\alpha=\{J,L,S,S_z\}$ and the coefficients  ${a_{\alpha
}^{(i)}}^{2}$, $b_{\alpha }^{2}$, ${c_{\alpha }^{(i)}}^{2}$, and
${d_{\alpha }^{(i)}}^{2}$ are defined as
\begin{eqnarray}\label{a1}
     {a_{\alpha }^{(1)}}^{2}(r,\rho,T)=r^{2}I_{L,S_{z}}(r,\rho,T),
 \end{eqnarray}
\begin{eqnarray}
     {a_{\alpha }^{(2)}}^{2}(r,\rho,T)=r^{2}[\beta I_{J-1,S_{z}}(r,\rho,T)
     +\gamma I_{J+1,S_{z}}(r,\rho,T)],
 \end{eqnarray}
\begin{eqnarray}
           {a_{\alpha }^{(3)}}^{2}(r,\rho,T)=r^{2}[\gamma I_{J-1,S_{z}}(r,\rho,T)
           +\beta I_{J+1,S_{z}}(r,\rho,T)],
      \end{eqnarray}
\begin{eqnarray}
     b_{\alpha }^{2}(r,\rho,T)=r^{2}[\beta _{23}I_{J-1,S_{z}}(r,\rho,T)
     -\beta _{23}I_{J+1,S_{z}}(r,\rho,T)],
 \end{eqnarray}
\begin{eqnarray}
         {c_{\alpha }^{(1)}}^{2}(r,\rho,T)=r^{2}\nu _{1}I_{L,S_{z}}(r,\rho,T),
      \end{eqnarray}
\begin{eqnarray}
        {c_{\alpha }^{(2)}}^{2}(r,\rho,T)=r^{2}[\eta _{2}I_{J-1,S_{z}}(r,\rho,T)
        +\nu _{2}I_{J+1,S_{z}}(r,\rho,T)],
 \end{eqnarray}
\begin{eqnarray}
       {c_{\alpha }^{(3)}}^{2}(r,\rho,T)=r^{2}[\eta _{3}I_{J-1,S_{z}}(r,\rho,T)
       +\nu _{3}I_{J+1,S_{z}}(r,\rho,T)],
 \end{eqnarray}
\begin{eqnarray}
     {d_{\alpha }^{(2)}}^{2}(r,\rho,T)=r^{2}[\xi _{2}I_{J-1,S_{z}}(r,\rho,T)
     +\lambda _{2}I_{J+1,S_{z}}(r,\rho,T)],
 \end{eqnarray}
\begin{eqnarray}\label{d2}
     {d_{\alpha }^{(3)}}^{2}(r,\rho,T)=r^{2}[\xi _{3}I_{J-1,S_{z}}(r,\rho,T)
     +\lambda _{3}I_{J+1,S_{z}}(r,\rho,T)],
 \end{eqnarray}
with
\begin{eqnarray}
          \beta =\frac{J+1}{2J+1},\ \gamma =\frac{J}{2J+1},\
          \beta _{23}=\frac{2J(J+1)}{2J+1},
 \end{eqnarray}
\begin{eqnarray}
       \nu _{1}=L(L+1),\ \nu _{2}=\frac{J^{2}(J+1)}{2J+1},\
       \nu _{3}=\frac{J^{3}+2J^{2}+3J+2}{2J+1},
      \end{eqnarray}
\begin{eqnarray}
     \eta _{2}=\frac{J(J^{2}+2J+1)}{2J+1},\ \eta _{3}=
     \frac{J(J^{2}+J+2)}{2J+1},
 \end{eqnarray}
\begin{eqnarray}
     \xi _{2}=\frac{J^{3}+2J^{2}+2J+1}{2J+1},\
     \xi _{3}=\frac{J(J^{2}+J+4)}{2J+1},
 \end{eqnarray}
\begin{eqnarray}
     \lambda _{2}=\frac{J(J^{2}+J+1)}{2J+1},\
     \lambda _{3}=\frac{J^{3}+2J^{2}+5J+4}{2J+1},
 \end{eqnarray}
and
\begin{eqnarray}\label{besselintegral}
       I_{J,S_{z}}(r,\rho,T)=\frac{1}{2\pi^{6}\rho^{2}}
       \int dk_{1}\ dk_{2}\ \overline{n}_{i}(k_m,T,B,\rho^{(i)})
        \overline{n}_{j}(k_m,T,B,\rho^{(j)})
       J_{J}^{2}(|k_2-k_1|r)\cdot
 \end{eqnarray}
In the above equation, $J_{J}(x)$ is the Bessel function.

Now, we minimize the two-body energy with respect to the variations
in the function $f_{\alpha}^{(i)}$ subject to the normalization
constraint,
\begin{eqnarray}
        \frac{1}{N}\sum_{ij}\langle ij\left| h_{S_{z}}^{2}
        -f^{2}(12)\right| ij\rangle _{a}=0.
 \end{eqnarray}
The minimization subject to the above normalization constraint leads
to the normalization of the two body wave function to unity
\cite{Owen}.
For the spin polarized hot neutron matter, the function
$h_{S_{z}}(r)$ is defined as follows,
\begin{eqnarray}
h_{S_{z}}(r)&=& \left\{\begin{array}{ll} \left[ 1-\left(
\frac{\gamma_i(r)}{\rho}\right) ^{2}\right] ^{-1/2} &;~~ S_{z}=\pm1,   \\ \\
1 &;~~ S_{z}= 0,
\end{array}
\right.
\end{eqnarray}
where
\begin{eqnarray}\label{intgamma}
       \gamma_i(r)=\frac{1}{2\pi^{2}}
       \int dk\  \overline{n}_{i}(k_m,T,B,\rho^{(i)})
       J_{0}(k r)k^2.
 \end{eqnarray}
From the minimization of the two-body cluster energy, we get a set
of coupled and uncoupled differential equations \cite{BordMod98}. By
solving these equations, we can obtain the correlation functions to
compute the two-body energy term, $E_2$.
As the final step, we calculate the free energy per particle, $F$,
to get different thermodynamic properties of spin polarized hot
neutron matter,
\begin{eqnarray}\label{free1}
       F(\rho,T,B)=E(\rho,T,B)-T S(\rho,T,B),
 \end{eqnarray}
where S is the entropy per particle,
\begin{eqnarray}\label{entropy}
S(\rho,T,B)&=&-\frac{1}{N}\sum_{i=+,-}\sum_{k}\{
{[1-\overline{n}_{i}(k,T,B,\rho^{(i)})]
\textrm{ln}[1-\overline{n}_{i}(k,T,B,\rho^{(i)})]}
\nonumber\\&+&{\overline{n}_{i}(k,T,B,\rho^{(i)}) \textrm{ln}[
\overline{n}_{i}(k,T,B,\rho^{(i)})]}\}.
\end{eqnarray}
It should be noted that in our calculations, we introduce the
effective masses, $m_{i}^{*}$, as variational parameters
\cite{Modarres3,Modarres5,Modarres7,Modarres8,Friedman}. We minimize
the free energy with respect to the variations in the effective
masses, and then we obtain the chemical potentials and the effective
masses of spin-up and spin-down neutrons at the equilibrium state.
In our approach, the effective mass depends on both density and
temperature but it is independent of the momentum. The effective
mass of a quasiparticle near the Fermi surface for the spin polarized
neutron matter at low temperatures is also the static physical
quantity of interest in the context of Landau Fermi liquid theory
\cite{Landau}.
%-------------------------------------------------------------------------
\section{RESULTS and DISCUSSION}\label{NLmatchingFFtex}

\begin{figure}
\includegraphics{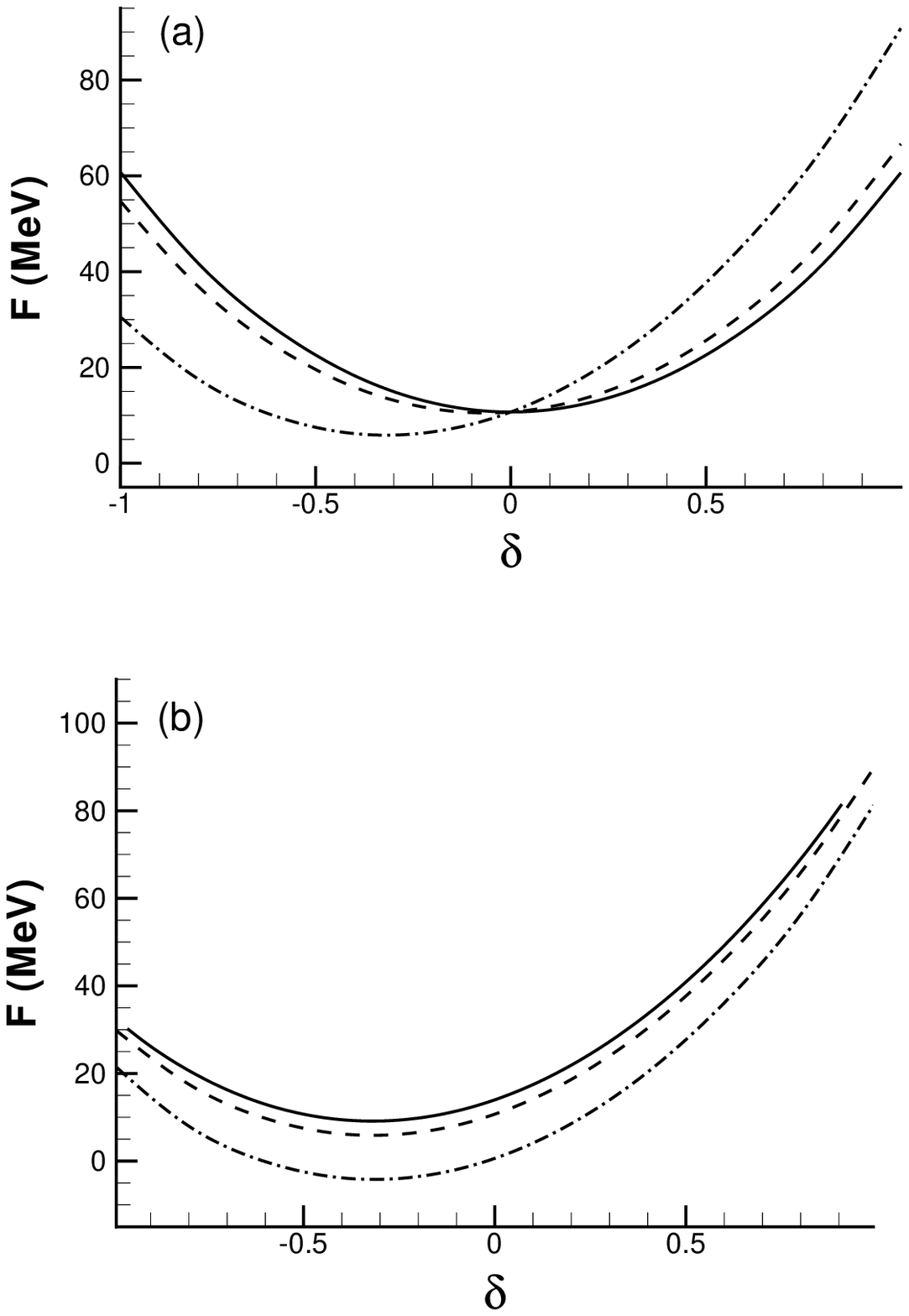}

\caption{\label{fig:1j} Free energy per particle as a function of
the spin polarization parameter $\delta$: (a) for the cases $B=0\ G$
(solid curve), $B=10^{18}\ G$ (dashed curve) and $B=5\times10^{18}\
G$ (dashdot curve) at the fixed values of the temperature, $T= 10\
MeV$, and the density, $\rho=0.2\ fm^{-3}$, (b) for the cases $T=0\
MeV$ (solid curve), $T=10\ MeV$ (dashed curve) and $T=20\ MeV$
(dashdot curve) at the fixed values of the magnetic field,
$B=5\times10^{18}\ G$, and the density, $\rho=0.2\ fm^{-3}$.}
\end{figure}

In Fig. \ref{fig:1j}, we present the free energy per particle of
spin polarized neutron matter versus the spin polarization parameter
$\delta$.
Fig. \ref{fig:1j}a shows that in the presence of magnetic field, the
free energy is not a symmetric function of spin polarization
parameter and the equilibrium configuration would experience a net
magnetization.
Clearly, the effects of magnetic fields below $B\sim10^{18}\ G$ are
almost insignificant, but by increasing the magnetic field, the
equilibrium value of the spin polarization parameter (i.e. the
polarization that minimizes the free energy) and the free energy
decrease, leading to a more stable system.
Fig. \ref{fig:1j}b indicates that the effect of temperature on the
fully spin polarized neutron matter is less than that of the unpolarized
one.

\begin{figure}
\includegraphics{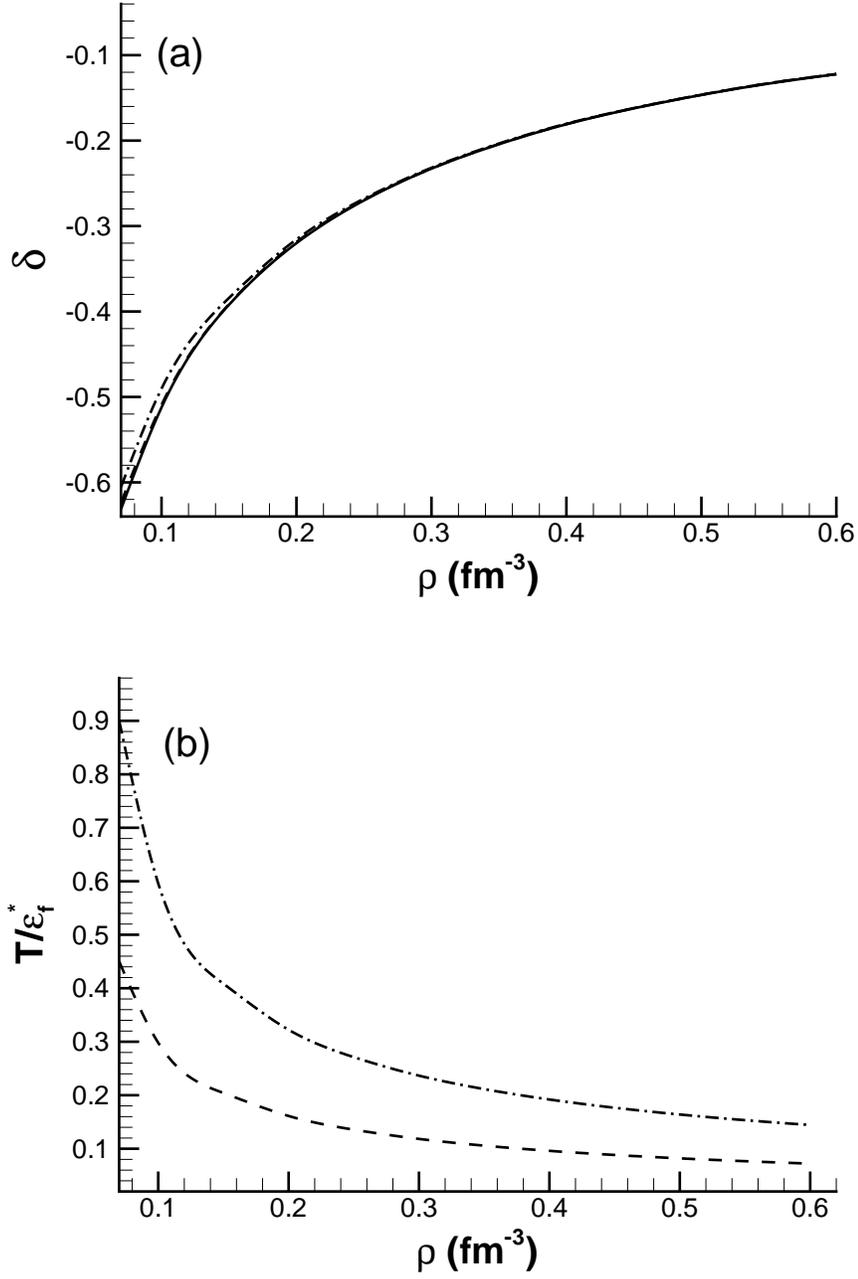}

\caption{\label{fig:2j} (a) Spin polarization parameter at the
equilibrium state as a function of the density $\rho$ for the cases
$T=0\ MeV$ (solid curve), $T=10\ MeV$ (dashed curve) and $T=20\ MeV$
(dashdot curve), and a fixed value of the magnetic field,
$B=5\times10^{18}\ G$. (b) Same as in the top panel but for the
ratio $T/\varepsilon_f^*$. }
\end{figure}

\begin{figure}
\includegraphics{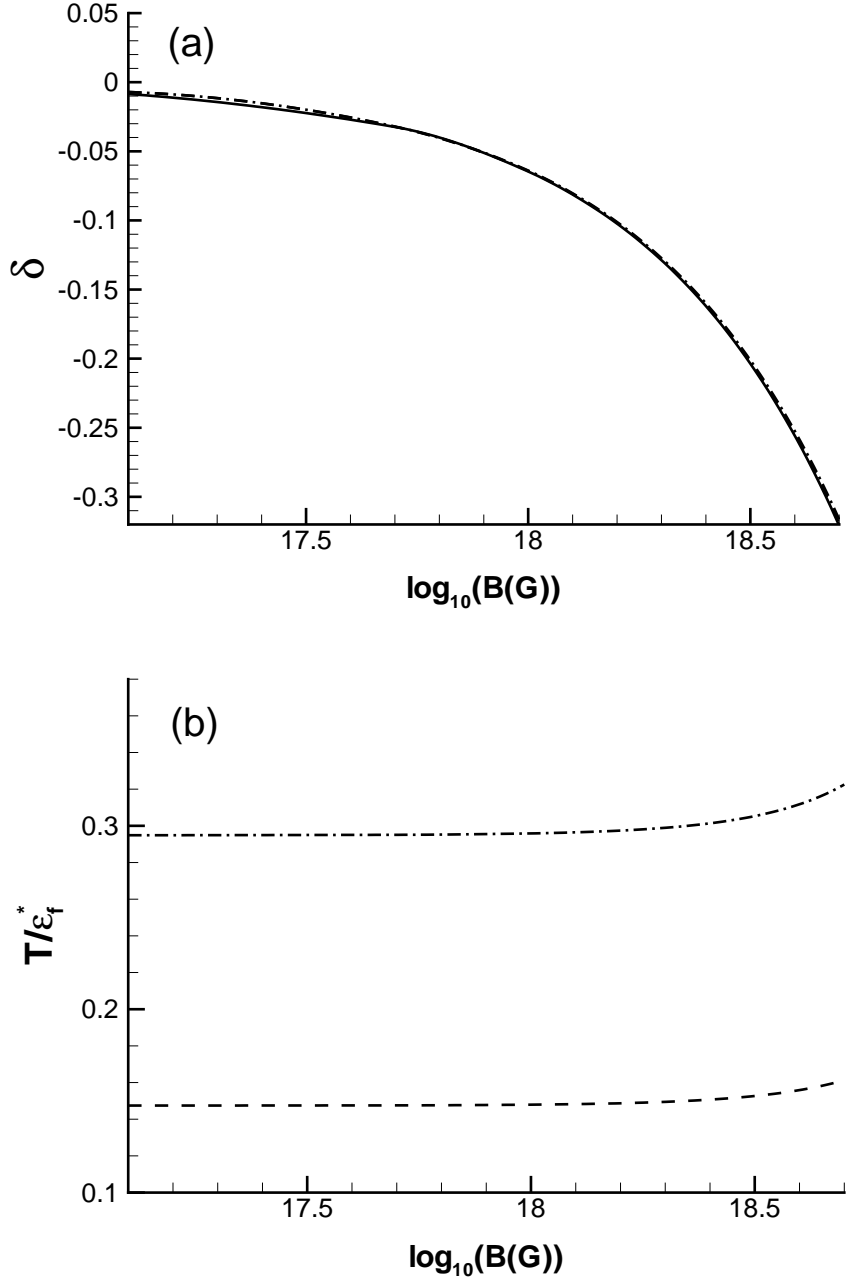}

\caption{\label{fig:3j} (a) Magnetic field dependence of the spin
polarization parameter $\delta$ at the equilibrium state for the
cases $T=0\ MeV$ (solid curve), $T=10\ MeV$ (dashed curve) and
$T=20\ MeV$ (dashdot curve), and a fixed value of the density,
$\rho=0.2\ fm^{-3}$. (b) Same as in the top panel but for the ratio
$T/\varepsilon_f^*$.}
\end{figure}

Fig. \ref{fig:2j} presents the equilibrium value of the spin
polarization parameter versus density $\rho$. Fig. \ref{fig:2j}a
shows that at low densities ($\rho \leq 0.2\ fm^{-3}$), the
magnitude of the spin polarization parameter decreases by increasing
the temperature. However, at higher densities, the related values of
the spin polarization parameter at different finite temperatures are
almost identical to the one for zero temperature.
This is due to smaller values of $T/\varepsilon_f^*$ at high
densities (Fig. \ref{fig:2j}b), in which $\varepsilon_f^*$ is
defined as follows,
\begin{eqnarray}\label{efffermi}
\varepsilon_f^*=\sum_{i=+,-}\frac{\rho^{(i)}}{\rho}\varepsilon_{fi}^*,
 \end{eqnarray}
with
\begin{eqnarray}
\varepsilon_{f+}^*=\frac{\hbar^{2}k_{F}^{(+)^{2}}}{2m}-\mu_n B,
 \end{eqnarray}
and
\begin{eqnarray}
\varepsilon_{f-}^*=\frac{\hbar^{2}k_{F}^{(-)^{2}}}{2m}+\mu_n B.
 \end{eqnarray}
In the above equations, $\varepsilon_{fi}^*$ and
$k_{F}^{(i)}=(6\pi^{2}\rho^{(i)})^{\frac{1}{3}}$ are the Fermi
energy and Fermi momentum of neutrons with spin projection $i$ in
the presence of the magnetic field. It is evident from Eq.
(\ref{efffermi}) that $\varepsilon_f^*$ gives an average of the
Fermi energy of magnetized neutron matter. Therefore, the ratio
$T/\varepsilon_f^*$ is a criterion for the fraction of particles
which are thermally excited \cite{Pathria}, and how much the system
is disordered. In Fig. \ref{fig:3j}, we show the spin polarization
parameter at the equilibrium state as a function of the magnetic
field $B$.
At each temperature, the magnitude of spin polarization parameter
grows by increasing the magnetic field (Fig. \ref{fig:3j}a).
We have found that at strong magnetic fields, the effect of finite
temperature is more significant because the ratio
$T/\varepsilon_f^*$ rises with the increase in the magnetic field
(Fig. \ref{fig:3j}b).

\begin{figure}
\includegraphics{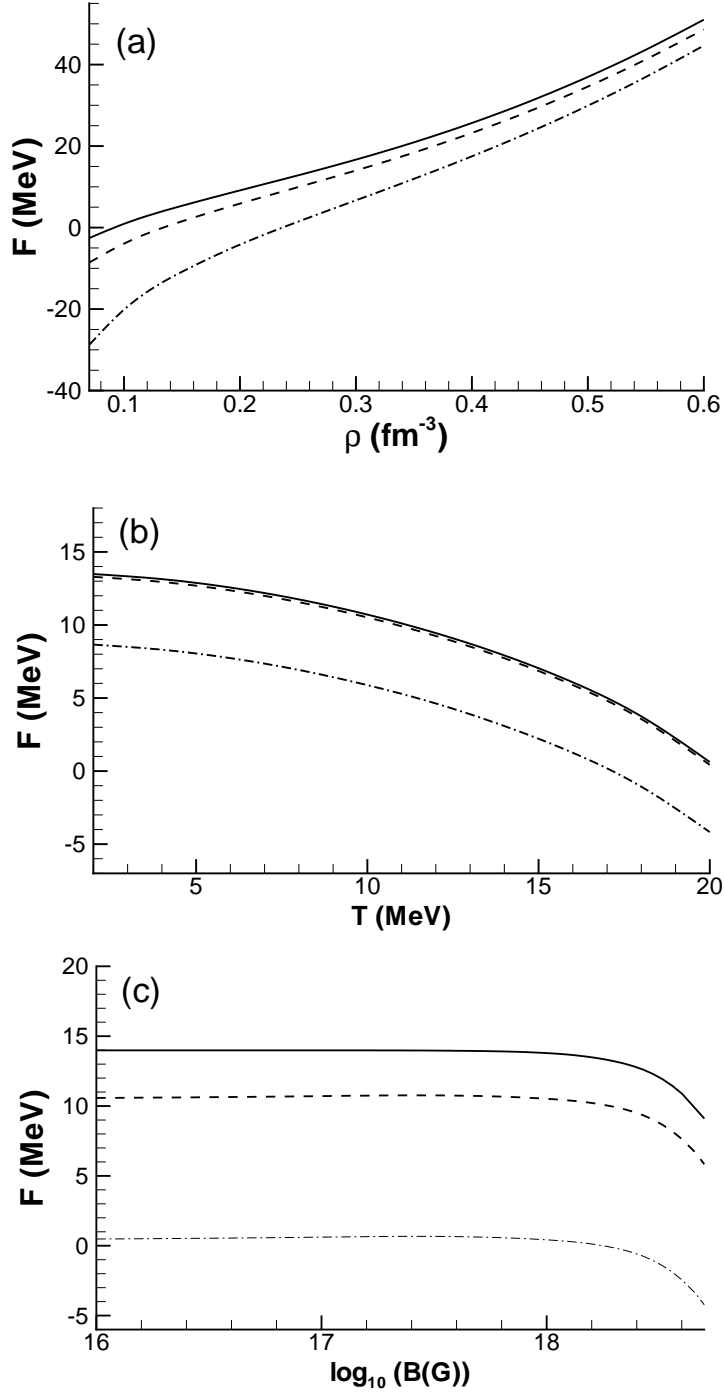}

\caption{\label{fig:4j} Free energy per particle at the equilibrium
state as a function of: (a) the density $\rho$ for the cases $T=0\
MeV$ (solid curve), $T=10\ MeV$ (dashed curve) and $T=20\ MeV$
(dashdot curve), and a fixed value of the magnetic field,
$B=5\times10^{18}\ G$, (b) the temperature $T$ for the cases $B=0\
G$ (solid curve), $B=10^{18}\ G$ (dashed curve) and
$B=5\times10^{18}\ G$ (dashdot curve), and a fixed value of the
density, $\rho=0.2\ fm^{-3}$, (c) the magnetic field $B$ for the
cases $T=0\ MeV$ (solid curve), $T=10\ MeV$ (dashed curve) and
$T=20\ MeV$ (dashdot curve), and a fixed value of the density,
$\rho=0.2\ fm^{-3}$.}
\end{figure}

The free energy per particle at the equilibrium value of the spin
polarization parameter is presented in Fig. \ref{fig:4j}.
It can be seen that at finite temperature, the free energy is an
increasing function of density (Fig. \ref{fig:4j}a).
At low densities, the rate of increase in the free energy varies by
increasing the density, but at high densities, this rate of increase
is nearly constant.
Moreover, the effect of temperature on the free energy is more
pronounced at low densities.
Fig. \ref{fig:4j}b shows that the free energy decreases by growing
the temperature nearly at the same rate for different magnetic
fields.
The free energy decreases by increasing the magnetic field (Fig.
\ref{fig:4j}c).
We can see that by increasing the magnetic field up to a value of
about $B \simeq 10^{18}\ G$, the free energy per particle slowly
decreases, and then it rapidly decreases for the magnetic fields
greater than this value. This indicates that, above $B \simeq
10^{18}\ G$, the effect of magnetic field on the free energy of the
spin polarized neutron matter becomes more important.

\begin{figure}
\includegraphics{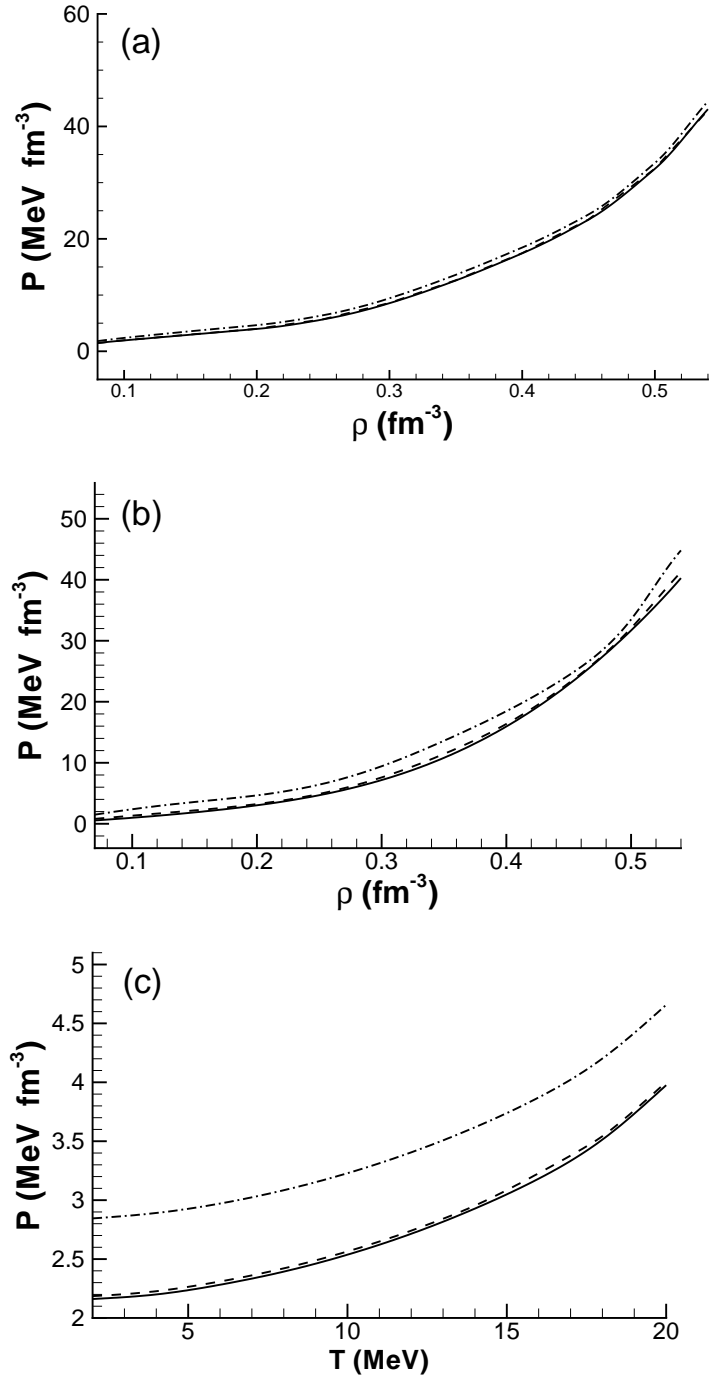}

\caption{\label{fig:5j} Pressure of spin polarized neutron matter as
a function of: (a) the density $\rho$ for the cases $B=0\ G$ (solid
curve), $B=10^{18}\ G$ (dashed curve) and $B=5\times10^{18}\ G$
(dashdot curve), and a fixed value of the temperature, $T= 20\ MeV$,
(b) the density $\rho$ for the cases $T=0\ MeV$ (solid curve),
$T=10\ MeV$ (dashed curve) and $T=20\ MeV$ (dashdot curve), and a
fixed value of the magnetic field, $B=5\times10^{18}\ G$, (c) the
temperature $T$ for the cases $B=0\ G$ (solid curve), $B=10^{18}\ G$
(dashed curve) and $B=5\times10^{18}\ G$ (dashdot curve), and a
fixed value of the density, $\rho=0.2\ fm^{-3}$.}
\end{figure}

From the free energy per particle of magnetized neutron matter, $F$,
we can obtain the corresponding pressure of neutron matter using the
following relation,
\begin{eqnarray}
      P(\rho,T,B)= \rho^{2}
      {\left(\frac{\partial F(\rho,T,B)}
      {\partial \rho} \right)
_{T,B}}.
 \end{eqnarray}
This equation of state (EoS) is plotted in Fig. \ref{fig:5j}.
For each value of the density, pressure increases by growing the
magnetic field (Fig. \ref{fig:5j}a).
This stiffening of the equation of state is due to the inclusion of
neutron anomalous magnetic moments.
From the astrophysical point of view, it should be noted that this
stiffening of the EoS leads to the larger value for the maximum mass
of neutron star \cite{Bordbar2006,Bordbar2009,Bordbar2011}.
At each density, the pressure at finite temperature is larger than
that of zero temperature (Fig. \ref{fig:5j}b). It means that the
equation of state of neutron matter becomes stiffer by increasing
the temperature.
Fig. \ref{fig:5j}c also shows that by increasing the temperature,
the pressure increases nearly at the same rate for different
magnetic fields.

\begin{figure}
\includegraphics{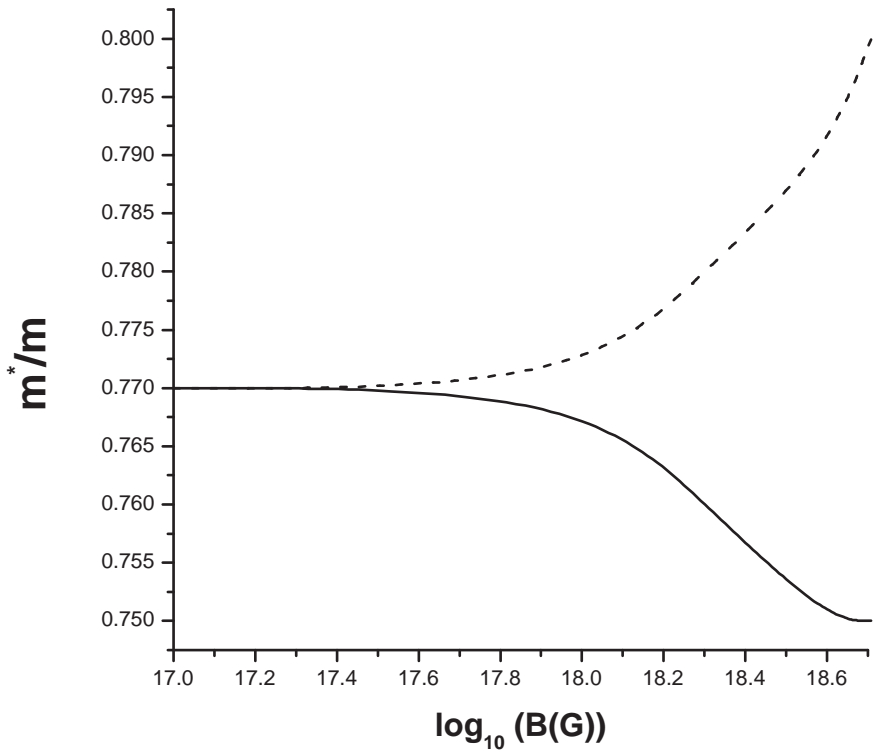}

\caption{\label{fig:6j} Magnetic field dependence of the effective
masses of spin-up (solid curve) and spin-down (dashed curve)
neutrons corresponding with the equilibrium state at the fixed
values of the temperature, $T= 10\ MeV$, and the density, $\rho=0.3\
fm^{-3}$. }
\end{figure}

Fig. \ref{fig:6j} shows the effective mass corresponding with the
equilibrium of the system for the spin-up and spin-down neutrons as
a function of the magnetic field $B$. At low magnetic fields, the
effective masses of spin-up and spin-down neutrons are nearly
identical because the effective masses of spin-up and spin-down
neutrons have the same values at $\delta \simeq 0$. Fig.
\ref{fig:6j} indicates that the effective mass of spin-up
(spin-down) neutrons decreases (increases) by increasing the
magnetic field in agreement with the results obtained in Ref.
\cite{Perez-Garcia2}. From the comparison of Fig. \ref{fig:3j}a and
\ref{fig:6j}, we can see that the shift in mass is due to the
polarization of neutron matter. For the maximum value of the
magnetic field considered in this work, i.e. $5 \times 10^{18}\ G$,
and at $T= 10\ MeV$ and $\rho=0.3\ fm^{-3}$, the equilibrium value
of the spin polarization parameter is about $\delta_{B_{max}}=-0.23$
and that corresponds to a mass shift of an amount of $\bigtriangleup
(m^*/m)\approx 0.02$ with respect to the unpolarized case.

%----------------------------------------------------------------------------------
\section{Summary and Concluding Remarks}
We have investigated the effect of strong magnetic fields on the
thermodynamic properties of spin polarized hot neutron matter
applying LOCV method and using $AV_{18}$ potential.
We have found that in the presence of a strong magnetic field, the
free energy is not a symmetric function of the spin polarization
parameter and the system is macroscopically magnetized.
By increasing both density and temperature, the magnitude of the
equilibrium value of the spin polarization parameter decreases.
At low magnetic fields, the free energy decreases very slowly by
increasing the magnetic field, but at stronger magnetic fields, the
free energy decreases rapidly with the increase in the magnetic
field.
It has been found that the equation of state becomes stiffer by
increasing the magnetic field.
This stiffening of the EoS leads to the larger value for the maximum
mass of neutron star.
%

%%%%%%%%%%%%%%%%%%%%%%%%%%%%%%%%%%%%%%%%%%%%%
\acknowledgements{ We wish to thank the Research
Institute for Astronomy and Astrophysics of Maragha and Shiraz University Research Council.}

%%%%%%%%%%%%%%%%%%%%%%%%%%%%%%%%%%%%%%%%%%%%%%%%%%%%%%%%%%%%%%%%%%%

%%%%%%%%%%%%%%%%%%%%%%%%%%%%%%%%%%%%%%%%%%%%%%%%%%%%%%%%%%%%%%%%%%%%%%%%%%%%%%%%%%%%%%%%%%%%%%%%%%%%%%
%%%%%%%%%%%%%%%%%%%%%%%%%%%%%%%%%%%%%%%%%%%%%%%%%%%%%%%%%%%%%%%%%%%%%%%%%%%%%%%%%%%%%%%%%%%%%%%%%%%%%%
\newpage
%%%%%%%%%%%%%%%%%%%%%%%%%%%%%%%%%%%%%%%%%%%%%%%%%%%%%%%%%%%%%%%%%%%%%%%%%%%%%%%%%%%%%%%%%%%%%%%%%%%%%%
\newpage

%%%%%%%%%%%%%%%%%%%%%%%%%%%%%%%%%%%%%%%%%%%%%%%%%%%%%%%%%%%%%%%%%%%%%%%%%%%%%%%%%%%%%%%%%%%%%
\newpage
%%%%%%%%%%%%%%%%%%%%%%%%%%%%%%%%%%%%%%%%%%%%%%%%%%%%%%%%%%%%%%%%%%%%%%%%%%%%%%%%%%%%%%%%%%%
\newpage
%%%%%%%%%%%%%%%%%%%%%%%%%%%%%%%%%%%%%%%%%%%%%%%%%%%%%%%%%%%%%%%%%%%%%%%%%%%%%%%%%%%%%%%%%%%
\newpage
%%%%%%%%%%%%%%%%%%%%%%%%%%%%%%%%%%%%%%%%%%%%%%%%%%%%%%%%%%%%%%%%%%%%%%%%%%%%%%%%%%%%%%%%%%%
%%%%%%%%%%%%%%%%%%%%%%%%%%%%%%%%%%%%%%%%%%%%%%%%%%%%%%%%%%%%%%%%%%%%%%%%%%%%%%%%%%%%%%%%%%%
\newpage
%%%%%%%%%%%%%%%%%%%%%%%%%%%%%%%%%%%%%%%%%%%%%%%%%%%%%%%%%%%%%%%%%%%%%%%%%%%%%%%%%%%%%%%%%%%

\end{document}